\documentstyle[psfig,longtable]{aipproc}

\begin{document}
\title{Strong Isospin Breaking in CP-even and CP-odd
$K\rightarrow\pi\pi$ Decays}
\author{C.E. Wolfe$^{*}$ and K. Maltman$^{**}$}
\address{$^*$Nuclear Theory Center, Indiana Univ., Bloomington, 
Indiana, USA\thanks{Supported by DOE grant \#DE-FG0287ER-40365} }
\address{$^{**}$Dept. Mathematics and Statistics, York Univ.,
4700 Keele St., Toronto, ON Canada, and CSSM, Univ. of Adelaide, 
Adelaide, SA Australia\thanks{Supported by the Natural Sciences
and Engineering Research Council of Canada} }
\maketitle
\begin{abstract}
Complete next-to-leading (chiral) order (NLO) expressions for
the strong isospin-breaking (IB) contributions in
$K\rightarrow\pi\pi$ are used to discuss (1) for CP-even,
the impact on the magnitude of the $\Delta I=1/2$ Rule, and (2)
for CP-odd, the strong IB correction, $\Omega_{st}$,
for the gluonic penguin contribution to $\epsilon^\prime /\epsilon$,
with particular emphasis on the strong low-energy
constant (LEC) and loop contributions, numerical values for which
are model-independent at NLO.
\end{abstract}
In the presence of IB, the standard isospin
decomposition of the $K^+\rightarrow\pi^+\pi^0$, 
$K^0\rightarrow\pi^+\pi^-,\pi^0\pi^0$ decay amplitudes, $A_{+0}$,
$A_{+-}$ and $A_{00}$, becomes~\cite{cdg}
\begin{eqnarray}
\label{isodecomp}
A_{00} &=&  \sqrt{1/3} A_0 
{\rm e}^{i\Phi_0}-[{\sqrt{2/3}}] A_2 {\rm e}^
{i\Phi_2}, \nonumber \\
A_{+-} &=& \sqrt{1/ 3} A_0 
{\rm e}^{i\Phi_0}+[{1/\sqrt{6}}] A_2 {\rm e}^
{i\Phi_2}, \nonumber \\
A_{+0} &=& [{\sqrt{3}/ 2}] A_2^\prime {\rm e}^{i\Phi_2^\prime}.
\end{eqnarray}
Ignoring $\Delta I=2$ electromagnetic (EM) channel coupling,
$\Phi_I$ are the $\pi\pi$ phases.
In general, EM- and strong-IB-induced
$\Delta I=5/2$ effects make
$\vert A_2^\prime\vert \not= \vert A_2\vert$.
$A_0$, $A_2$ can be chosen real
in the absence of CP violation.

Since $\vert A_0\vert\sim 22 \vert A_2\vert$,
IB ``leakage'' of the large
octet ($I=0$) amplitude into the 
$\Delta I=3/2$ ($I=2$) amplitude can be numerically significant.
EM leakage has been
computed to NLO in Ref.~\cite{cdg} (see also \cite{eimnp00}); 
we compute the NLO strong leakage
for both CP-even and gluonic-penguin-mediated CP-odd transitions.
Strong cancellation
between gluonic penguin ($O_6$) 
and electroweak penguin contributions make
Standard Model predictions for $\epsilon^\prime /\epsilon$
sensitive to the latter~\cite{burasrev}.

At leading (chiral) order (LO), the ratio, $\delta A_2/A_0$, with
$\delta A_2$ the strong octet leakage contribution to
$A_2$, is unambiguous. 
The IC part of $A_2$,
$A^{IC}_2$, is {\it decreased} by $\left[ \Omega_{st}\right]_{LO}=13\%$
once strong IB is included.
The LO 
$O_6$ suppression in $\epsilon^\prime /\epsilon$
is $1-\Omega_{st}$.  Recent analyses of $\epsilon^\prime /\epsilon$
employ $\Omega_{st}=0.25\pm 0.08$~\cite{burasrev}. The difference 
reflects estimates of NLO effects mediated by an intermediate
$\eta^\prime$ through the $K^0\rightarrow\pi^0\eta^\prime$
transition~\cite{dghtbg}.  This effect, in
conventional ChPT (involving only the
$\pi$, $K$ and $\eta$ as explicit degrees of freedom), represents one
contribution to the CP-even (CP-odd) 
NLO weak LEC's, $E^+_k$ ($E^-_k$)~\cite{kmw90}, but
does not exhaust such contributions.

Let us fix notation.  At LO, the low-energy representations of the dominant
CP-even ($+$) octet operator and CP-odd ($-$) gluonic penguin operator
are~\cite{kmw90}
\begin{equation}
c^{\pm}Tr\left[ \lambda^{\pm}\partial_\mu U^\dagger
\partial^\mu U\right]\ ,
\label{LO}\end{equation}
where $\lambda^+=\lambda_6$, $\lambda^-=\lambda_7$,
and $U=exp\left( i{\lambda}\cdot{\pi}\right)$, as usual.
A complete set of NLO contributions is obtained by evaluating the
sum of (i) graphs with a single LO weak vertex (proportional
to $c^\pm$) and one of the external
legs dressed by a single strong NLO vertex (proportional to one
of the NLO strong LEC's, $L_k$) 
(``strong LEC contributions''); (ii) 1-loop graphs
involving a single LO weak vertex and strong vertices only of LO
(``loop contributions''); and
(iii) tree graphs involving a single NLO weak vertex~\cite{kmw90} 
(``NLO weak LEC
contributions'').  Only the sum of all three classes of 
contribution is renormalization-scale-independent
and physically meaningful.
From Eq.~(\ref{LO}), 
it is immediately obvious that LO CP-even and CP-odd $K^0$ 
decay vertices are related through the substitution 
$c^+ \leftrightarrow {\rm i}\, c^-$.  
The loop-plus-strong-LEC part of the NLO contribution
to the {\it ratio}
$\delta A_2/A_0$ is, therefore, the same for the CP-even and CP-odd cases.
Because, moreover, the LO LEC's, $c^\pm$, cancel in the
ratio, this contribution is,
though scale-dependent, {\it model}-independent.

The strong LEC contributions to $\delta A_2/A_0$
have recently been discussed in Ref.~\cite{emnp99}.  They include an
$\eta^\prime$-mediated $\pi^0$-$\eta$ mixing term
proportional to $L_7$ which is, however, empirically, almost completely
cancelled by an accompanying $L_8$
contribution~\cite{emnp99}.  The loop contributions
have been evaluated in Ref.~\cite{cwkmk2pi}.  Since
the full set of $E^+_k$ ($E^-_k$) cannot be
determined empirically at present, one is forced to use
model values.  The models used are discussed below.

\begin{table}
\caption{Strong octet and EM IB leakage contributions in
units of $10^{-6}\ {\rm MeV}$. The isospin-conserving (IC) and
LO IB fits
yield $A_2 = A_2^\prime = -2.1\times 10^{-5}\ {\rm MeV}$
and $-2.4\times 10^{-5}$ MeV, respectively. }\label{table1}
\begin{tabular}{lcc}
\hline
Source\ \ \qquad\qquad&$\delta A_2$&
$\delta A_2^\prime$ \\ 
\hline
$(8)\qquad$&$(-1.56\pm 0.63) +(0.42\pm 0.05){\rm i}$&
\qquad $(-1.56\pm 0.63)+(0.42\pm 0.05){\rm i}$ \\
$(EM)$\qquad&$(-1.27\pm 0.40)
-(1.28\pm 0.02){\rm i}$&
\qquad $(0.70\pm 0.73)-(0.07\pm 0.04){\rm i}$ \\
\hline
\end{tabular}
\end{table}
For the CP-even case, the leakage contributions,
$\delta A_2=\delta A_2^\prime$, are
given in Table 1~\cite{cwkmk2pi}.  The corresponding EM 
contributions~\cite{cdg} are shown for comparison.
The errors reflect uncertainties in the estimates of the
NLO LEC's (see Ref.~\cite{cwkmk2pi} for details, 
and references to the models employed for the
$E^+_k$).  Denoting the ratio of fitted LO
$27$-plet to octet weak LEC's obtained neglecting,
or including, IB by $r_{IC}$, or $r_{IB}$, respectively,
we find 
$R_{IB}\equiv r_{IB}/r_{IC}=0.963\pm 0.029\pm 0.010\pm 0.034$.
The errors correspond, respectively, to model dependence of the $E^+_k$,
uncertainties in
$B_0(m_d-m_u)$, and the EM uncertainties of Refs.~\cite{cdg}.  
The deviation from $1$
is significantly smaller than at LO (where $R_{IB}=0.870$).
The true $\Delta I=1/2$ rule
enhancement, $r_{IB}$, can thus be
approximated by $r_{IC}$ to an accuracy of better than $10\%$.
The $\Delta I=5/2$ contribution (dominantly EM in
character~\cite{cwkmk2pi}), leads to
${A_2}/{ A_2^\prime}=1.094\pm 0.039\not= 1$,
and significantly exacerbates the phase discrepancy problem
for the neutral $K$ decays~\cite{cwkmk2pi}.

For the CP-odd case, 
$\Omega_{st}=\omega \, {{\rm Im}\, \delta A_2}/\, {{\rm Im}\, A_0}$ 
($\omega= {\rm Re}\, A_0/{\rm Re}\, A_2\simeq 22$).  ${\rm Im}\, A_0$ is
associated with $O_6$.
At LO, $\left[ \Omega_{st}\right]_{LO}=0.13$.  At NLO
\begin{equation}
{\frac{\Omega_{st}}{[\Omega_{st}]_{LO}}}=\left[
1+{\frac{{\rm Im}\, \delta A_2^{(NLO;ND)}}
{{\rm Im}\, \delta A_2^{(LO)}}}-{\frac{{\rm Im}\, A_0^{(NLO;ND)}}
{{\rm Im}\, A_0^{(LO)}}}\right]\equiv
\left[ 1+R_2-R_0\right]\ .
\end{equation}
The superscript $(NLO;ND)$ indicates the sum of non-dispersion
NLO contributions (involving NLO weak and strong LEC's
and the non-dispersive parts of loop graphs).  
Neither the NLO $I=0$ isospin-conserving
nor NLO $I=2$ IB leakage $E^-_k$ combinations are known.
The NLO dispersive contributions create phases consistent
with Watson's theorem.
Although the positive $I=0$ phases correspond to
attractive FSI, the $E^-_k$ terms may, nonetheless,
make ${\rm Im}\, A_0$ smaller at NLO than the LO
(see comments on Ref.~\cite{pichpallente} 
in Ref.~\cite{burasetal}
for a related discussion).
If, however, NLO effects {\it do} enhance $Im\, A_0$ 
(decreasing the level of $O_6$-$O_8$ cancellation and
increasing $\epsilon^\prime /\epsilon$) 
$\Omega_{st}$ will be simultaneously suppressed,
further amplifying this increase.
The known NLO contributions (loops and strong
LEC terms) give contributions $-0.24 (-0.31)$ to $R_2$
and $-0.02 (+0.42)$ to $R_0$,
at scale $\mu =m_\eta (m_\rho )$~\cite{cwkmee}.  
For the (model-dependent) $E^-_k$ contributions,
we considered the possibility of using 3 models:
the weak deformation model (WDM)~\cite{wdm},
the chiral quark model (ChQM)~\cite{chiqm}, and the
scalar saturation model of Ref.~\cite{gvshazi} (SSM).  Ref.~\cite{gvshazi}
was the first to argue explicitly that NLO weak LEC
contributions other than those mediated by the $\eta^\prime$
might be numerically important.  A large 
negative contribution to $R_2-R_0$ 
was obtained which, if correct,
would significantly enhance the $O_6$ contribution
to $\epsilon^\prime /\epsilon$.  In contrast, the two earlier
models produce positive, or near-zero, $E_k^-$
contributions to $R_2-R_0$.
Since there is
a problem with the SSM estimate\begin{footnote}
{The model values, $\hat{E}_k^-$, of the
$E_k^-$
are obtained by (1) approximating the renormalized
4-quark operator $O_6$ as the product
of renormalized scalar densities (the factorization approximation)
and (2) dropping (divergent) seagull terms in the low-energy
representation of this product.  Two types of contribution to
$\hat{E}_k^-$ result, $\left[ \hat{E}_k^-\right]_1$ 
(proportional to the {\it strong}
$6^{th}$ order LEC's), and $\left[ \hat{E}_k^-\right]_2$
(proportional to a product of 2 $4^{th}$ order {\it strong} LEC's).  
As an example, 
$\left[ \hat{E}_1^-\right]_2 = 16\sqrt{2}G_FV_{us}^*V_{ud}\, Im\, (C_6)
B_0^2\, L_8^2$~\cite{gvshazi}.  
In Ref.~\cite{gvshazi} it is assumed that the finite part
of this expression is 
proportional to the square, $[L_8^r]^2$, of the finite part
of $L_8$.  This, however, is incorrect.
In dimensional regularization,
the Laurent expansion of $L_8$ begins at $O[1/(d-4)]$;
the finite part of $L_8^2$ is thus actually
$[L_8^r]^2 +5L_8^{(-1)}/384\pi^2$, with $L_8^{(-1)}$ 
the coefficient of the $O[d-4]$ term in the $L_8$ Laurent expansion.
$L_8^{(-1)}$ enters physical processes
beginning at $6^{th}$ order in the chiral expansion, and
hence is on the same footing as the other $6^{th}$ order strong
LEC's appearing in $\left[ \hat{E}_k^-\right]_1$
and retained in Ref.~\cite{gvshazi}.  Since, it is
inconsistent to drop the $L_k^{(-1)}$, and the SSM provides no means
of estimating their values, the numerical estimates given in
Ref.~\cite{gvshazi} are incomplete, and cannot be used.}\end{footnote}
we have used only the WDM and ChQM. 
Taking the WDM model $E_k^-$
to correspond to a scale $\mu\sim m_\rho$, we find,
combining with the strong LEC and loop contributions
given above, 
$1+R_2-R_0=0.27$ in the WDM and $0.64$ in the ChQM.
Taking the more microscopic ChQM to provide a central value, and
the difference between the WDM and ChQM to provide a
{\it minimal} measure of model dependence in the
theoretical result, we 
obtain, at NLO,
$\Omega_{st}=0.08\pm 0.05$, significantly
smaller than both the conventionally employed 
value $0.25\pm 0.08$, and that, $0.16\pm 0.03$,
obtained in the estimate of Ref.~\cite{emnp99},
based on the strong
LEC contributions only.
The central value above, combined
with conventional central values for the $B$-factors,
leads to a $\sim 50\%$ increase in the 
predicted value for $\epsilon^\prime /\epsilon$.  
To be conservative, we would propose using this
lower central value 
with an even larger error estimate, in all future calculations
of Standard Model values for $\epsilon^\prime /\epsilon$. 

\end{document}